\documentclass[aps,twocolumn,showpacs,floatfix,amsmath,amssymb,prb,superscriptaddress,]{revtex4-2}
\usepackage[hypertexnames=false]{hyperref}
\usepackage[version=4]{mhchem}
\usepackage[utf8]{inputenc}
\usepackage{graphicx}
\usepackage{xcolor}
\usepackage[normalem]{ulem}
\makeatletter

\begin{document} 
\newcommand{\be}{\begin{equation}}
\newcommand{\ee}{  \end{equation}}
\newcommand{\ba}{\begin{eqnarray}}
\newcommand{\ea}{  \end{eqnarray}}
\newcommand{\ve}{\varepsilon}
\newcommand{\joel}[1]{\textcolor{red}{#1}}
\newcommand{\albertin}[1]{\textcolor{blue}{#1}}
\newcommand{\albertout}[1]{\textcolor{blue}{\sout{#1}}}

\title{Magnetoresistivity in the Antiferromagnetic Hubbard Model}

\author{Joel Bobadilla} \affiliation{Universidad de Buenos Aires, 
     Facultad de Ciencias Exactas y Naturales, Departamento de F\'{i}sica. Buenos Aires, Argentina}

\author{Marcelo J. Rozenberg} \affiliation{Universit\'e Paris-Saclay, CNRS Laboratoire de Physique des Solides, 91405, Orsay, France}
\affiliation{Universit\'e Paris-Cité, CNRS
Integrative Neuroscience and Cognition Center, 75006, Paris, France}

\author{Alberto Camjayi} \email{alberto@df.uba.ar} \affiliation{Universidad de Buenos Aires, 
  Ciclo B\'asico Com\'un. Buenos Aires, Argentina} \affiliation{CONICET-Universidad de Buenos Aires,
  Instituto de F\'isica de Buenos Aires (IFIBA). Buenos Aires, Argentina}

\date{\today}

\begin{abstract}
We investigate the magnetotransport properties of the half-filled antiferromagnetic (AF) one-band Hubbard model under an external magnetic field using the single-site dynamical mean-field approximation (DMFT). Particular attention is paid to the mechanisms driving the magnetoresistivity behavior. We analyze the dependence of magnetoresistivity on temperature and the strength of the applied magnetic field, providing insights into the interplay between magnetic field induced fluctuations and transport properties in AF systems. 
\end{abstract}



\maketitle


\section{\label{Intro} Introduction}

Magnetoresistivity (MR) refers to the change in the electrical resistivity of a material subjected to an external magnetic field. The MR ratio, usually expressed as a percentage, is defined as

\be
\mathrm{MR}(h)=\frac{\rho(h)-\rho(0)}{\rho(0)}
\ee
where $\rho(h)$ and $\rho(0)$ are the resistivity measured in the presence and
absence of a magnetic field $h$, respectively. 

The study of MR has a long and rich history in condensed matter physics. Since its discovery in 1856 by William Thomson (Lord Kelvin), MR has found multiple applications across a wide range of technologies, including magnetic storage, sensing, control, and computation. In simple metals, MR is positive, meaning the resistivity increases with the applied field. However, in certain transition metal alloys and ferromagnets, negative MR has also been observed. Depending on the magnitude and physical origin, several classes of MR have been identified: ordinary MR (OMR), giant MR (GMR), colossal MR (CMR), and tunneling MR (TMR).
 
In nonmagnetic metals, the Lorentz force deflects the trajectories of charge carriers, increasing the probability of scattering events and thereby enhancing resistivity. This phenomenon, known as OMR, is typically small (on the order of $\sim 5\%$), positive, and follows a quadratic $h^2$ dependence at low fields and temperatures. In 2007, Albert Fert and Peter Gr\"unberg were awarded the Nobel Prize in Physics for their discovery of GMR~\cite{butler,tsymbal,grunberg}. Unlike OMR, GMR is a negative and significantly larger effect (typically 10--50\% or more), arising from multilayered ferromagnetic–nonmagnetic heterostructures, such as Fe/Cr and Co/Cu multilayers~\cite{grunberg}. In GMR devices, when the magnetic moments of the ferromagnetic layers are aligned parallel, scattering is weak and resistivity is low; when they are antiparallel, scattering is strong and resistivity increases. Since an external magnetic field tends to align the moments into a parallel configuration, the resistivity decreases with field, resulting in a negative magnetoresistance. 

Colossal magnetoresistance (CMR), observed in manganites with a perovskite structure, often exceeds GMR by an order of magnitude, reaching MR values over 100\%~\cite{ramirez,tokura}. The underlying mechanisms behind CMR are complex and believed to involve double-exchange interactions, Jahn–Teller distortions, and phase separation phenomena~\cite{dagoto}. Prototypical compounds exhibiting CMR are the strongly correlated 
manganites, such as $\mathrm{La}_{1-x}\mathrm{Sr}_x\mathrm{MnO}_3$, which were intensely studied during the 1990s~\cite{ramirez,tokura}.

Tunneling MR (TMR) occurs in magnetic–nonmagnetic
multilayer heterostructures, typically with nanometric insulating barriers. The tunneling probability depends on the relative alignment of the magnetic moments across the barrier~\cite{butler}: parallel alignment enhances tunneling and reduces resistance, whereas antiparallel alignment suppresses tunneling and increases resistance. Typical materials exhibiting TMR effects include Co/Al$_2$O$_3$/Co~\cite{butler} and perovskite manganite-based tunnel junctions~\cite{dagoto}.

In all of the above phenomena, ferromagnetism plays a crucial role. Antiferromagnetic (AF) analogues, such as AF GMR and AF TMR, have been proposed theoretically in so-called AF spin valves and AF tunnel junctions,
however experimental realization of these AF counterparts remains challenging~\cite{baltz}. 

Nevertheless, AF materials are increasingly recognized as promising candidates for spintronic applications. Their inherent advantages over ferromagnetic systems include insensitivity to external magnetic fields, ultrafast spin dynamics, and higher device density~\cite{jungwirth,daldin}. Furthermore, AF materials are abundant and diverse, encompassing insulators, semiconductors, metals, and even superconductors.
The discovery of the so-called Néel-order spin-orbit torques~\cite{zelezny} has paved the way for utilizing AF materials as active elements in spintronic devices
~\cite{jungwirth,daldin}.
However, despite these promising features, significant challenges remain: manipulating and detecting the magnetic state of antiferromagnets is considerably more difficult than in ferromagnets, due to the absence of a net magnetization and the typically weak coupling to external probes.

To explore the fundamental mechanisms underlying magnetoresistivity in correlated antiferromagnets, we turn to one of the most widely studied and paradigmatic models in condensed matter theory: the one-band Hubbard model. Originally introduced by J. Hubbard in the 1960s to explain the emergence of magnetism in narrow-band transition metals such as iron~\cite{hubbard}, the Hubbard model has since become a central framework for understanding strong electronic correlations. Its relevance grew significantly after the discovery of high-temperature superconductivity in the cuprates, whose parent compounds are antiferromagnetic Mott insulators~\cite{imada}. Along similar lines, in the 1990s, manganites exhibiting colossal magnetoresistance (CMR) were recognized as strongly correlated materials~\cite{imada,ramirez,dagoto}, and high-quality thin films enabled a resurgence of experimental interest in field-tuned metal-insulator transitions. 

In this context, the Hubbard model provides a minimal yet powerful theoretical setting to investigate the interplay between magnetism, electronic correlations, and transport phenomena. While dynamical mean-field theory (DMFT) has been extensively applied to study the Hubbard model, especially in its single-site formulation~\cite{RMP13}, the magnetoresistivity behavior in the antiferromagnetic phase has received surprisingly little attention. Here, we aim to fill this gap by studying magnetotransport in the half-filled antiferromagnetic Hubbard model under an external magnetic field, using single-site DMFT on an infinite dimensional hypercubic bipartite lattice.

Our results reveal a basic mechanism for bulk magnetoresistivity, driven by sublattice-resolved fluctuations — in analogy with, but distinct from, GMR — and provide a solid benchmark for future studies using more realistic material-specific approaches such as LDA+DMFT~\cite{kotliar}. Throughout the paper, we critically compare our theoretical findings with experimental observations reported in correlated systems, aiming to bridge the gap between minimal models and material complexity.

\section{\label{description} Theoretical description}

As outlined in the Introduction, we explore the magnetotransport properties of the one-band Hubbard model.
At half-filling, this model is known to exhibit an antiferromagnetic (AF) ground state on a bipartite lattice~\cite{RMP13}. While its magnetic and spectral properties have been extensively studied over the years~\cite{RMP13,Pruschke2003,Zhu2017,Zitko2021}, the behavior of magnetoresistivity (MR) has received little attention. A notable exception is the study by Li \textit{et al.}~\cite{Li2018}, who investigated a one-dimensional Hubbard chain with periodically modulated interactions and found an enhanced MR,
highlighting the importance of lattice bipartition.

Here, we solve the half-filled Hubbard model on an infinite-dimensional bipartite hypercubic lattice using single-site dynamical mean-field theory (DMFT)~\cite{RMP13}, and study its behavior under a Zeeman-like magnetic field $h$ \footnote{We focus on the Zeeman coupling to the magnetic field and neglect orbital effects (via Peierls phases in the hopping terms), since our goal is to study the mechanisms driving MR in correlated materials. The most spectacular consequence of orbital effects is the appearance of the famous Hofstadter’s butterfly, whose physics requires unattainable high magnetic fields to be observed in solids ~\cite{Acheche2017}. This regime is, however, accessible in artificial structures such as cold-atom lattices and moiré superlattices. At experimentally realizable field strengths, orbital coupling instead leads to quantum oscillations, whose interplay with strong electronic correlations gives rise to highly intricate phenomena that remain the subject of active research ~\cite{Vucicevic2021,Yang2024}}. Our aim is to uncover the basic mechanisms responsible for MR in the AF phase and to establish a theoretical benchmark for future material-specific investigations.

\subsection{\label{model} The model}

To account for the bipartite lattice structure, we introduce the operators $a^\dagger$ and $b^\dagger$, which act on the two interpenetrating sublattices A and B, respectively. The Hamiltonian of the system is expressed as,
\begin{multline}
H = -t\sum_{\langle ij \rangle, \sigma}\left( a^\dagger_{i \sigma} b_{j \sigma} + b^\dagger_{i \sigma} a_{j \sigma} \right)+\\+ U\sum_i n_{i\uparrow}n_{i\downarrow} - \mu \sum_{i, \sigma} n_{i \sigma} - h \sum_{i} S^z_i 
\label{H} 
\end{multline}
where $a^\dagger_{i \sigma}$ and $b^\dagger_{i \sigma}$ ($a_{i \sigma}$ and $b_{i \sigma}$) are the creation (annihilation) operators for an electron with spin $\sigma$ on site $i$ of the corresponding sublattice. Here, $t$ is the hopping amplitude, $\langle ij \rangle$ indicates the summation over nearest-neighbor sites, $U$ is the on-site Coulomb repulsion, $\mu$ is the chemical potential, $n_{i \sigma}$ is the number operator for electrons with spin $\sigma$ on site $i$, and $h$ represents the strength of the external magnetic field applied along the z-direction.

By Fourier transforming the kinetic energy term in (\ref{H}), we get,
\be
H_\mathrm{kin} = \sum_{\sigma} \sum_{\mathbf{k} \in \mathrm{MBZ}}
 \Psi^\dagger_{\mathbf{k},\sigma}
\left(\begin{matrix}
        0 & \ve_\mathbf{k} \\
        \ve_\mathbf{k} & 0 
       \end{matrix} \right) \Psi_{\mathbf{k},\sigma},
\ee
where the summation is restricted to the magnetic Brillouin zone (MBZ), $\ve_\mathbf{k}$ denotes the energy dispersion, and $\Psi^\dagger_{\mathbf{k},\sigma} = \left(a^\dagger_{\mathbf{k} \sigma}\; b^\dagger_{\mathbf{k} \sigma}\right)$ is the spinor composed of the creation operators for the two sublattices. From this expression, the matrix Green's function of the system can be written as, 
\be 
\textbf{G}_{\sigma}(i\omega_n) = \sum_{\mathbf{k} \in \mathrm{MBZ}} \left( \begin{matrix}
                             \zeta_{A\sigma} & -\ve_\mathbf{k} \\
                             -\ve_\mathbf{k} & \zeta_{B\sigma} 
                              \end{matrix} \right)^{-1} \label{inv_G},
\ee                   
where $\zeta_{\alpha \sigma} = i \omega_n + \mu - \sigma h-\Sigma_{\alpha \sigma}(i \omega_n)$ with $\alpha = A,B$. Here, $\Sigma_{\alpha \sigma}(i \omega_n)$ represent the local self-energies on each sublattice, evaluated at Matsubara frequencies. 

\subsection{\label{dmft} DMFT solution of the model}

The model is solved using single-site dynamical mean-field theory (DMFT) \cite{RMP13} which maps a lattice model onto a quantum impurity model coupled to a non-interacting bath of electrons via a self-consistency condition. This method relies on the observation that in infinite dimensions, the self-energy $\Sigma(\textbf{k},i\omega_n)$ becomes $\textbf{k}$ independent \cite{Muller89, Metzner1989}, making a single-site treatment with only temporal fluctuations exact in this limit.

\subsubsection{\label{Self-consistency} Self-consistency condition}

Given the bipartite lattice structure of our system, two distinct impurity problems must be posed to obtain the DMFT equations. The self-consistency condition is expressed by, 
\be
\mathcal{G}_{0\alpha\sigma}^{-1}(i\omega_{n})=i\omega_{n}+\mu - \sigma h -\Delta_{\bar{\alpha}\sigma} \label{self-consistency},
\ee
where $\alpha= A,B$, $\bar{\alpha} = B,A$, and $\mathcal{G}_{0\alpha\sigma}$ is the Weiss field for each sublattice. Since we consider only nearest-neighbor hopping processes, the hybridization function $\Delta_{\bar{\alpha}\sigma}$ corresponds to electrons in the neighboring sublattice for each equation. 

The hybridization functions are given by, 
\be
\Delta_{\alpha \sigma}(i \omega_n) = i \omega_n + \mu - \sigma h -\Sigma_{\alpha \sigma}(i \omega_n) 
- G^{-1}_{\alpha \sigma}(i \omega_n) \label{delta},
\ee
with, \be
G_{\alpha \sigma}(i \omega_n) = \zeta_{\bar{\alpha}\sigma} 
\int_{-\infty}^{\infty} \mathrm{d}\ve \frac{D(\ve)}{\zeta_{A\sigma}\zeta_{B\sigma}-\ve^2}, \label{G}
\ee
the local Green's functions for each sublattice. Here, $D(\ve)=e^{-\ve^2/2 t^2}/\sqrt{2\pi t^2}$ is the density of states (DOS) for the hypercubic lattice. This expression arises from the integration of Eq. (\ref{inv_G}) over momentum space (or equivalently over energy).

\subsubsection{\label{AF} Antiferromagnetic (\textit{h} = 0) solution}

Usually, the model described in Sec.~\ref{model} is solved for $h=0$, in which case the system exhibits Néel Antiferromagnetic order, with A and B sublattices magnetized in opposite directions. Under these conditions, the Néel sublattice-spin symmetry imposes the additional relation $\Delta_{A\sigma}=\Delta_{B\bar{\sigma}}$ (the same is true for the Green's functions and self-energies), reducing the self-consistency relation (\ref{self-consistency}) to a single impurity problem, 
\be
\mathcal{G}_{0\sigma}^{-1}(i\omega_{n})=i\omega_{n}+\mu-\Delta_{\bar{\sigma}}. \label{AF-self-consistency}
\ee
It is sufficient to solve for both spin projections on one sublattice, as the complementary sublattice solution can be obtained by symmetry.

\subsubsection{\label{general} General solution}

To compute the magnetoresistance of the system, we need to solve the model in the presence of an external, uniform, and static magnetic field $h$. This field breaks the spin symmetry of the system, and the general bipartite case must be considered. At first glance, this may seem incompatible with the single-site DMFT framework, since the bipartite structure suggests an inherently two-site problem. However, the key insight is that we can still solve two single-site impurity problems — one for each sublattice — provided they are coupled self-consistently through the hybridization function (Eq.~(\ref{self-consistency})).

In this scenario, the Green’s function becomes a $4 \times 4$ block matrix in the combined spin and sublattice basis:
\be 
\mathbf{G}_{\mathbf{k}}(i\omega_n) = \left( \begin{matrix}
\zeta_{A\uparrow} & -\varepsilon_\mathbf{k} & 0 & 0 \\
-\varepsilon_\mathbf{k} & \zeta_{B\uparrow} & 0 & 0 \\
0 & 0 & \zeta_{A\downarrow} & -\varepsilon_\mathbf{k} \\
0 & 0 & -\varepsilon_\mathbf{k} & \zeta_{B\downarrow}
\end{matrix} \right)^{-1}. \label{inv_fullG}
\ee
The local Green's functions for each spin and sublattice projection are obtained by summing over momentum (or equivalently integrating over energy) the corresponding diagonal elements of $\mathbf{G}_{\mathbf{k}}(i\omega_n)$, leading to Eq.~(\ref{G}).

Although DMFT assumes a local self-energy, the bipartite structure of the lattice and the restriction to nearest-neighbor hopping ensure that each impurity site is dynamically coupled to its complementary sublattice. The off-diagonal structure of the full Green’s function mixes the self-energies of the two sublattices implicitly, enabling a fully self-consistent single-site DMFT treatment that captures the essential inter-sublattice feedback without violating the locality assumption.

The proposed solution scheme is as follows: we initialize the loop by providing an initial guess for the hybridization function of one sublattice, say $\Delta_{A\sigma}$, and for the self-energy of the complementary sublattice, $\Sigma_{B\sigma}$. The quantum impurity problem associated with $\Delta_{A\sigma}$ is then solved to obtain the corresponding self-energy $\Sigma_{A\sigma}$. Using this and the initial $\Sigma_{B\sigma}$, we compute $G_{B\sigma}$ via Eq.~(\ref{G}), and update $\Delta_{B\sigma}$ using Eq.~(\ref{delta}). This process is then repeated symmetrically, updating each sublattice iteratively until convergence is reached.

We have verified that this scheme, when applied at $h = 0$ without explicitly imposing the Néel sublattice-spin symmetry $\Delta_{A\sigma} = \Delta_{B\bar{\sigma}}$, reproduces the results obtained using the method described in Sec.~\ref{AF}.

\subsection{\label{Numerical_implementation} Numerical implementation}

The model was solved for $U=1.7$, which places the system in the weakly correlated regime~\cite{Trastoy2020}, with $t=0.5$ setting the energy scale throughout this work. We considered the half-filled case for which $\mu=U/2$. The DMFT is implemented by means of the continuous time diagramatic Quantum Monte Carlo (CT-QMC) method \cite{Haule2007}, a robust and versatile impurity solver which allows us to accurately compute the Green's functions and self-energies in the Matsubara frequency domain. Spectra are obtained via analytic continuation performed by the maximum entropy method (MaxEnt)~\cite{Levy2017Maxent}. The DC conductivity was computed using the Kubo formula as derived in Appendix~\ref{conduc}.

\section{Results and Discussion} \label{results}

In this section, we present the magnetization, magnetotransport, and spectral properties obtained for the model,
under various temperature and magnetic field conditions. Particular attention is given to the magnetoresistivity (MR) and its dependence on thermal agitation and external magnetic fields, as these provide crucial insights into the transition between antiferromagnetic (AF) insulating and paramagnetic (PM) metallic phases. MR is a sensitive probe of the interplay between electronic correlations and magnetic order, and our results highlight its critical role in understanding the system's behavior. To facilitate the discussion, we divide the results into two parts: temperature dependence and magnetic field strength dependence.

\subsection{\label{temperature_dependence} Temperature Dependence}

\begin{figure}[ht]
    \centering
    \includegraphics[width=\columnwidth]{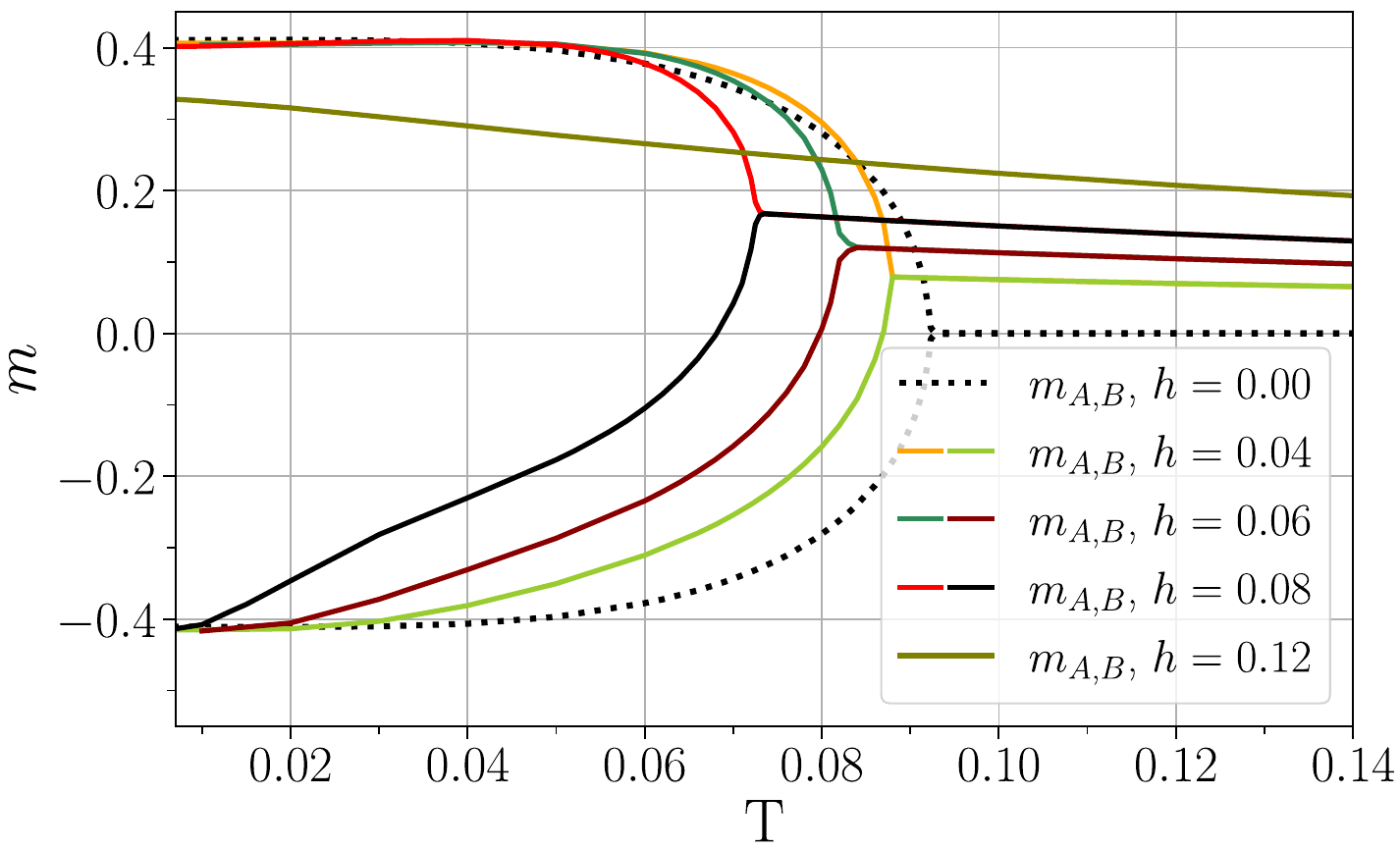}
    \caption{Temperature dependence of the magnetization \(m_A\) and \(m_B\) for sublattices \(A\) and \(B\) under different magnetic field values \(h\). The curves highlight the suppression of AF order with increasing temperature and the asymmetric alignment of spins due to the external magnetic field.}
    \label{fig:magnetization}
\end{figure}

We begin by analyzing the results for the magnetization of the two sublattices, $m_A$ and $m_B$, as a function of temperature. These results are shown in Figure~\ref{fig:magnetization}. The dotted lines represent the magnetization curves in the absence of an applied magnetic field ($h=0$). In this case, thermal agitation competes with AF coupling, eventually destroying the magnetic order at the Néel temperature $T_N \sim 0.092$. When $h\neq0$, the situation changes significantly.

The application of a non-zero external magnetic field 
breaks the sublattice-spin symmetry of the system. As the temperature increases, the magnetization of the antiparallel sublattice $m_B$ decreases more rapidly than that of the parallel sublattice $m_A$. At a specific temperature, $m_B$ converges with $m_A$, signaling a transition to a PM phase with a non-zero net magnetization.  If the magnetic field strength exceeds a critical threshold $h^*$, the AF order is completely suppressed, and the system remains an externally polarized paramagnet at all temperatures. This behavior is shown for $h=0.12$ in Figure~\ref{fig:magnetization}.

From an experimental standpoint, similar behavior has been observed in strongly correlated systems. The partial frustration of the AF order that we report at intermediate field strengths has also been identified in \ce{V2O3} samples \cite{Trastoy2020}. Regarding the existence of a critical field that fully suppresses the AF order, this phenomenon has been reported in various systems, including the heavy-fermion compounds \ce{CePtIn4} \cite{Das2019} and \ce{YbRh2Si2} \cite{Knapp2025}, thin films of \ce{V5S8} \cite{Hardy2016}, and the bilayer ruthenate Ti-doped \ce{Ca3Ru2O7} \cite{Zhu2016}.

\begin{figure*}[ht]
    \centering
    \includegraphics[width=\textwidth]{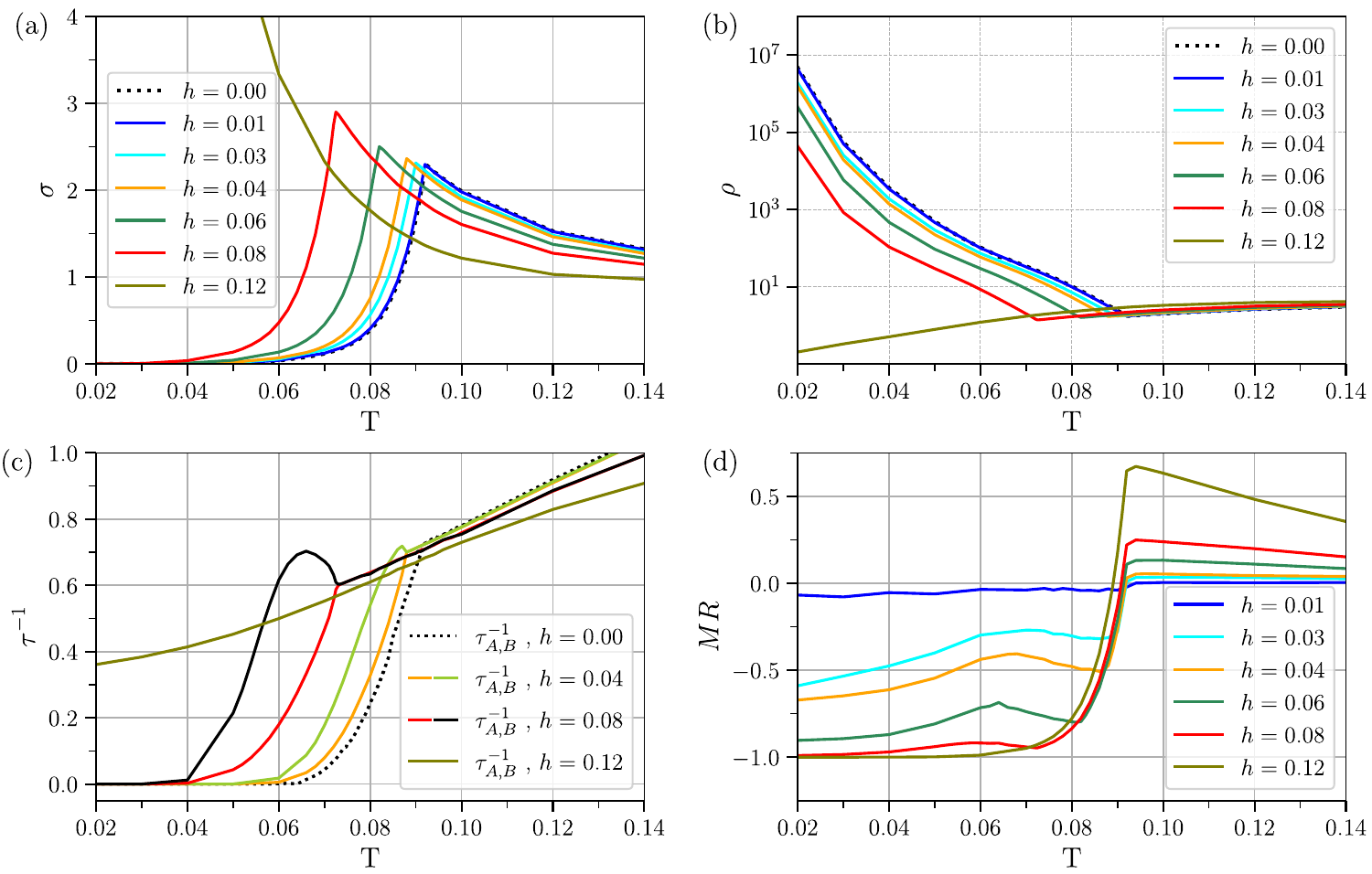}
    \caption{
        Temperature dependence of magnetotransport properties for various magnetic field strengths \(h\):
        (a) Conductivity \(\sigma(T)\),
        (b) Resistivity \(\rho(T)\) (logarithmic scale),
        (c) Scattering rates \(\tau^{-1}_A\) and \(\tau^{-1}_B\), and
        (d) Magnetoresistance (MR). The plots highlight the effect of the magnetic field on the system's transport properties across the temperature range studied.
    }
    \label{fig:transport_T}
\end{figure*}

\begin{figure}[ht] 
    \centering 
    \includegraphics[width=\columnwidth]{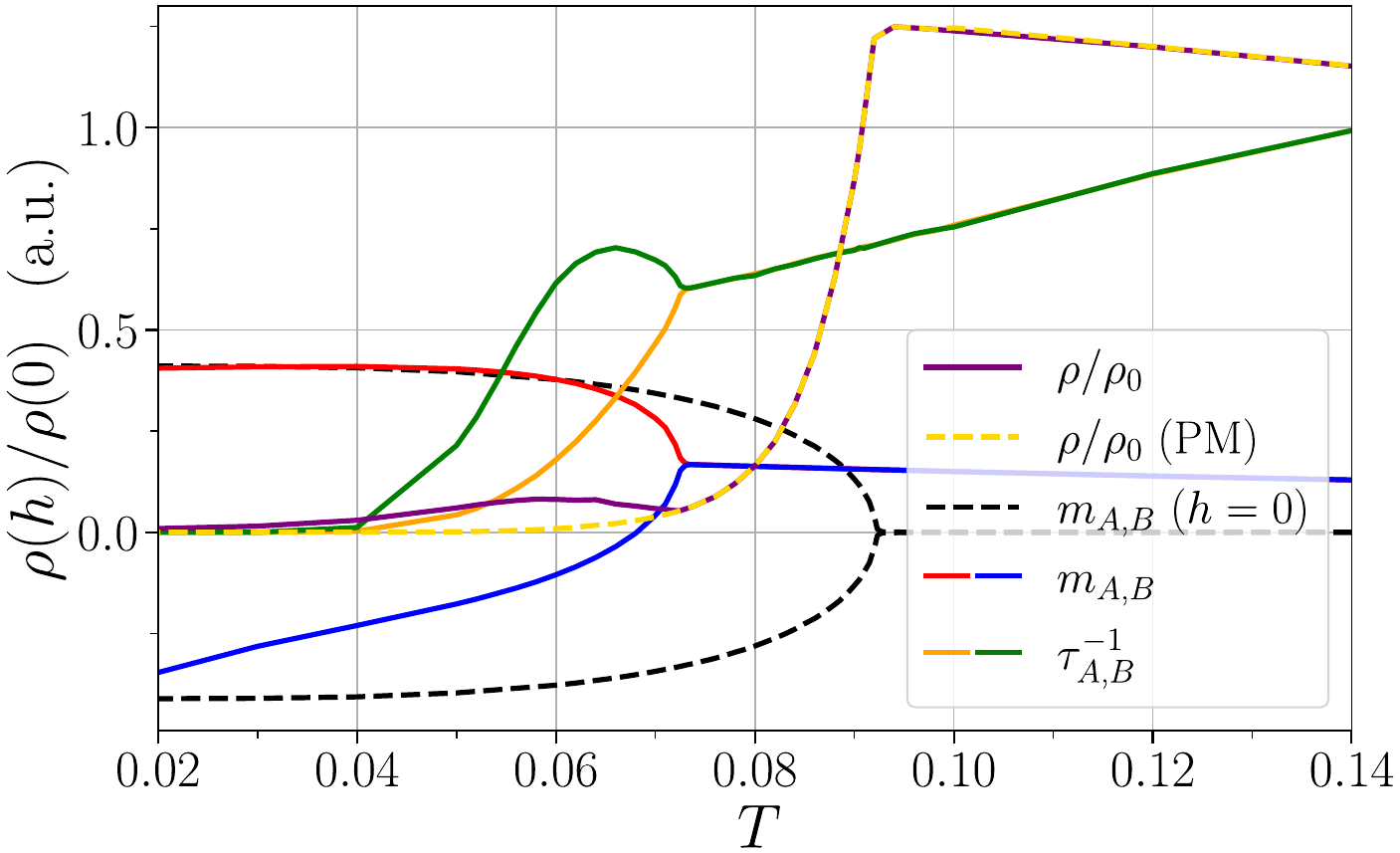} 
    \caption{Normalized resistivity $\rho(h)/\rho(0) = MR + 1$ obtained for the full AF solution (solid line) and the enforced PM state (dashed line) for $h=0.08$, plotted alongside the magnetization $m_A$ and $m_B$, with and without applied field, and the scattering rates $\tau^{-1}_A$ and $\tau^{-1}_B$. All quantities are plotted in arbitrary units (a.  u.) for clarity.} \label{fig:mag_rho_scatt_h0.08}
    \end{figure}

\begin{figure*}[ht]
    \centering
    \includegraphics[width=\textwidth]{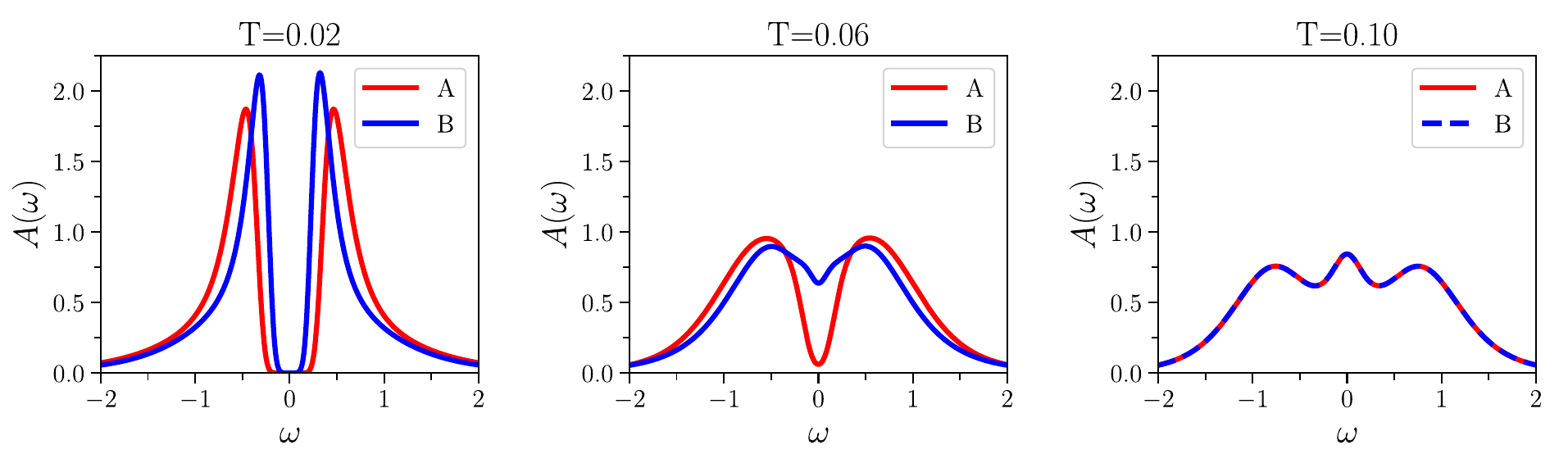}
    \caption{
        Spectral function \(A(\omega)\) as a function of real frequency \(\omega\) for \(h=0.08\), at various temperatures. The red and blue curves represent the spectral weight evolution for the parallel (A) and antiparallel (B) sublattices, respectively. As temperature increases, the gap for the B sublattice (blue) fills up faster than for the A sublattice (red), reflecting the asymmetry induced by the external magnetic field.
    }
    \label{fig:DOS_T}
\end{figure*}

Figure~\ref{fig:transport_T} presents the temperature dependence of the system's transport properties. Panels (a) and (b) show the conductivity $\sigma(T)$ and resistivity $\rho(T)$, respectively. As the temperature decreases, the system transitions from a PM metallic phase to an AF insulating phase. However, for sufficiently strong magnetic fields, such as $h=0.12$, the metallic behavior persists, as the AF order is completely suppressed. This highlights the dominant role of magnetic order in driving the metal-insulator 
transition in the one-band Hubbard model — a behavior also observed in correlated materials such as \ce{V2O3}~\cite{Trastoy2020} and light rare-earth nickelates~\cite{Ramadoss2016}. 

Panels (c) and (d) of figure~\ref{fig:transport_T} show the scattering rates, or inverse lifetimes, $\tau^{-1}$, of the two sublattices and the magnetoresistance (MR), respectively. The inverse lifetime $\tau^{-1}$ quantifies the rate at which quasiparticles lose coherence due to scattering processes \cite{BruusFlensberg}, with
$\tau^{-1} \propto -\Im (G^{-1}(\omega \to 0))$.

In panel (c), the scattering rates increase significantly in the AF phase when a non-zero magnetic field is applied. Furthermore, the scattering rates for the two sublattices, $A$ and $B$, split, with the sublattice $B$ experiencing stronger fluctuations. This behavior arises from the competition between the AF coupling and the applied field, which induces larger quantum fluctuations in the antiparallel sublattice ($B$). These fluctuations are further enhanced by thermal agitation. As the temperature decreases, the scattering rates for both sublattices decline, eventually converging to zero in the insulating phase. At high temperatures in the metallic PM phase, the application of the external field stabilizes the magnetic order, thereby reducing the scattering rate.

Interestingly, anomalous transport behavior associated with magnetic-field-induced suppression of AF order has also been reported in other correlated models. In particular, studies of the periodic Anderson model using DMFT have shown that a strong Zeeman field can destabilize AF order and drive the system into a non-Fermi-liquid metallic phase characterized by enhanced scattering and incoherent quasiparticles~\cite{Amaricci2008}.

Although, as in GMR systems, the transition from an antiparallel to a parallel magnetic configuration reduces electrical resistivity, our system is governed by sublattice-dependent rather than spin-dependent scattering dynamics. Thermal agitation and the external magnetic field reconfigure the sublattice magnetizations, altering scattering pathways, and modifying transport properties. In this case, the resistivity decrease results from a field-induced insulator-to-metal transition — a wavefunction localization transition — rather than from spin-dependent scattering mechanisms characteristic of GMR systems. 

Panel (d) highlights the MR dependence on temperature and its ability to capture all of these characteristics. At high temperatures, the MR is positive and gently decreasing, consistent with the behavior expected in non-magnetic metallic systems. However, as the temperature approaches $T_N\sim0.092$, the MR undergoes an abrupt change, decreasing sharply and becoming negative. This behavior occurs in the temperature range where the system remains in the metallic PM phase due to the suppression of the AF order by the external field. Essentially the same mechanism driving the MR sign reversal reported in AF \ce{V2O3} samples \cite{Trastoy2020}. 

As the temperature further decreases, the MR reaches a local minimum at the temperature where the transition to the AF insulating phase occurs. This minimum corresponds to the onset of magnetic order, where thermal agitation is sufficiently suppressed to allow the AF phase to stabilize. Below this point, the MR increases smoothly due to the splitting of the scattering rates, which is more pronounced in the antiparallel sublattice and leads to a rapid increase in resistivity. However, as the thermal agitation continues to fade, the MR growth decelerates and eventually reverses, decreasing again.

Notably, for $h=0.12$, the MR does not exhibit this final characteristic: the field suppresses AF order entirely, and no transition to the insulating state occurs. Figure~\ref{fig:mag_rho_scatt_h0.08} illustrates this behavior for $h = 0.08$, comparing $\rho(h)/\rho(0)$ from the full AF solution (solid) and from an enforced PM solution (dashed). Presenting $\rho(h)/\rho(0)$ instead of MR allows for simultaneous visualization of $m_A$, $m_B$, and scattering rates on the same scale.

A strikingly similar MR profile was reported in the mixed-valent vanadate \ce{PbV6O11} (Fig. 4 in~\cite{Maignan2010}), although in that case the magnetic transition is from a ferromagnetic to a PM phase, and the MR effect is attributed to tunneling across stacking faults. Other systems with temperature-dependent MR sign reversal include Ge films~\cite{Li2024}, bulk samples of \ce{Nd_{1-x}Sr_xNiO2} ($x = 0.2$, $0.4$)~\cite{Li2020}, and thin films of AF \ce{Mn3Pt}~\cite{Mukherjee2021}. In Ge and nickelates, the origin of the MR reversal remains unclear, while in Mn$_3$Pt it has been linked to spin reorientation effects. This diversity highlights the variety of physical mechanisms that can govern MR behavior in correlated electron systems.

To conclude this section, we analyze the temperature evolution of the spectral function $A(\omega)$, shown in Fig.~\ref{fig:DOS_T} for $h = 0.08$. At high temperatures, both sublattices exhibit a quasiparticle peak at the Fermi level ($\omega = 0$), characteristic of the metallic PM phase. As temperature decreases, the external field lifts sublattice degeneracy, and a gap begins to form asymmetrically. In sublattice $A$ (parallel to the field), the gap opens more rapidly due to alignment with the external polarization, which reduces fluctuations and favors AF order. In contrast, sublattice $B$ experiences stronger fluctuations and a delayed gap opening due to the competing effects of AF coupling and the external field. At low temperatures, both gaps become fully developed, signaling the emergence of the AF insulating state. This asymmetry reflects the field-driven imbalance between sublattices and underscores the intricate interplay of magnetic order, quantum fluctuations, and correlation-driven localization.

\subsection{\label{Magnetic_Field_Strength_Dependence} Magnetic Field Strength Dependence}

\begin{figure}[ht]
    \centering
    \includegraphics[width=\columnwidth]{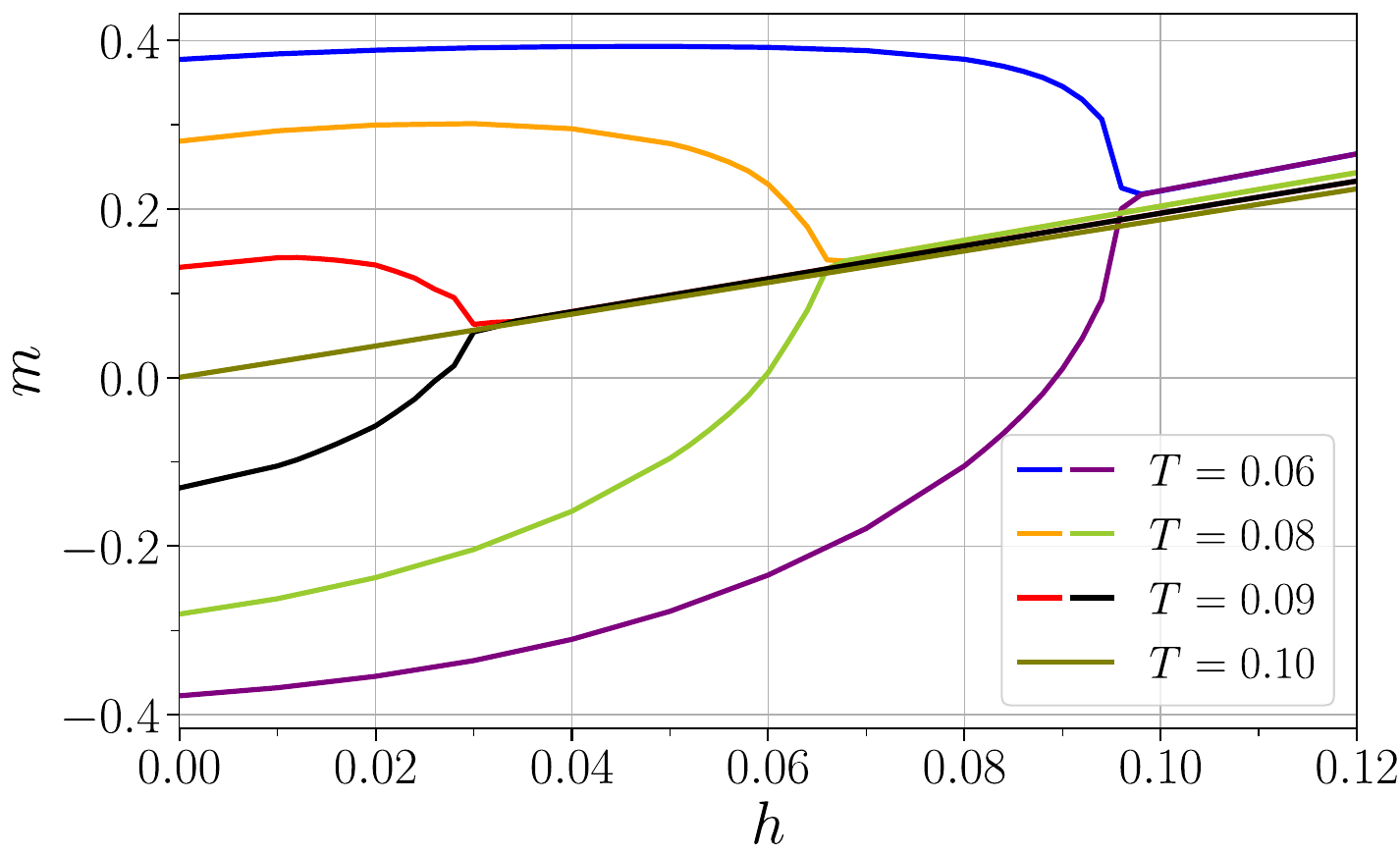}
    \caption{       
    Magnetization \(m_A\) (upper set of curves) and \(m_B\) (lower set) as a function of magnetic field strength \(h\) for various fixed temperatures.}
    \label{fig:magnetization_h}
\end{figure}

\begin{figure*}[ht]
    \centering
    \includegraphics[width=\textwidth]{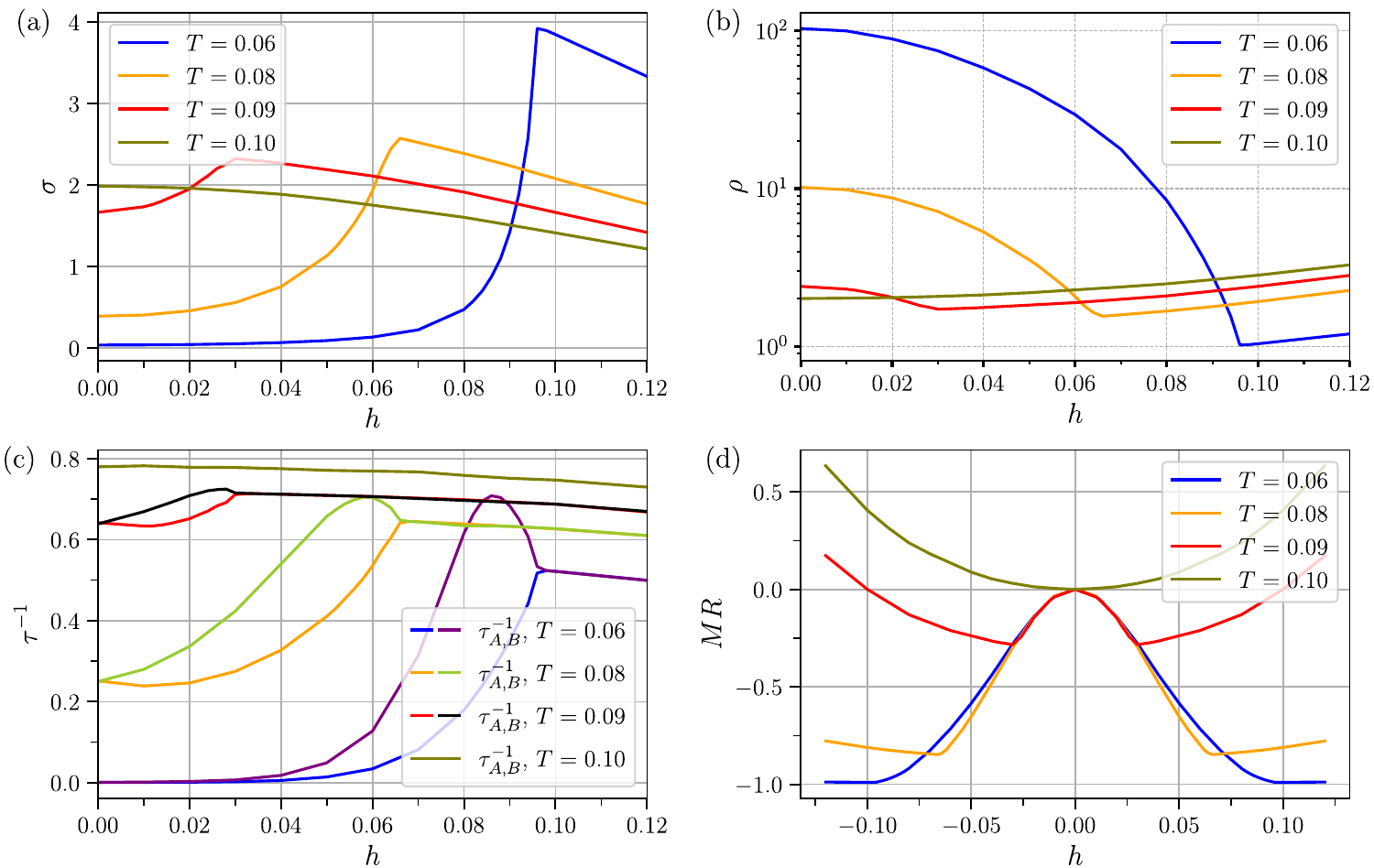}
    \caption{
        Transport properties of the system as a function of the magnetic field strength h at different temperatures: (a) Conductivity $\sigma(h)$: Highlights the transition from a bad insulating to a metallic behavior at higher fields. 
        (b) Resistivity $\rho(h)$: plotted on a logarithmic scale. The change in the sign of the derivative indicates the field-driven insulator-to-metal transition.
        (c) Scattering rates $\tau^{-1}_A$ and $\tau^{-1}_B$: display sublattice asymmetry induced by the competition between Zeeman energy and AF correlations
        (d) Magnetoresistance (MR): Shows how at low fields the MR follows a $h^2$ law, transitioning from negative values at low temperatures to positive values at higher temperatures.
    }
    \label{fig:transport_H}
\end{figure*}

\begin{figure*}[ht]
    \centering
    \includegraphics[width=\textwidth]{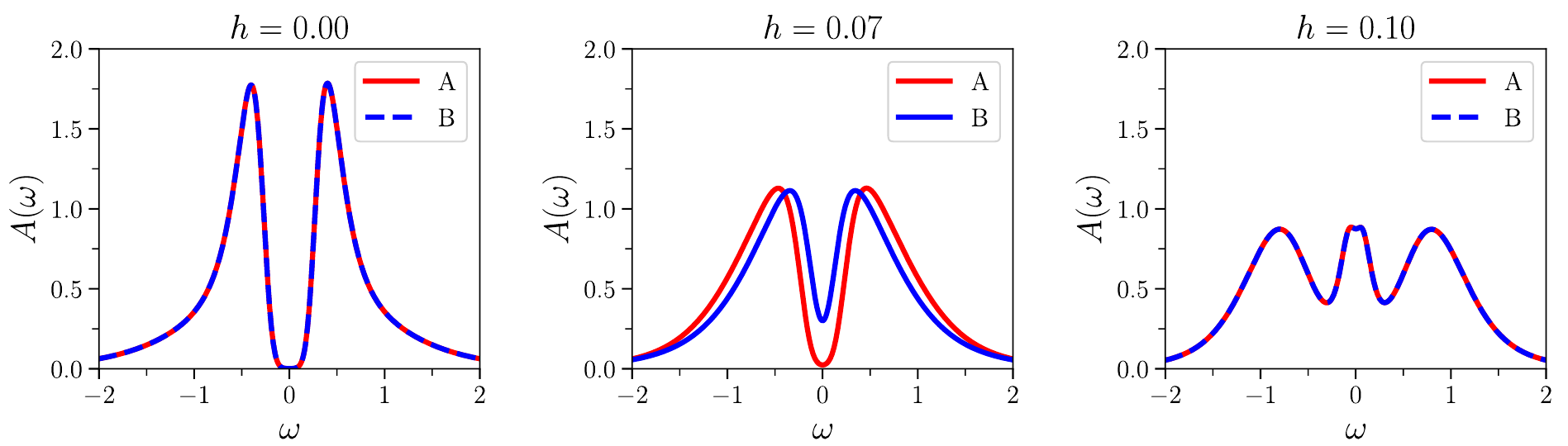}
    \caption{
        Spectral function \(A(\omega)\) as a function of real frequency \(\omega\) for \(T=0.06\), at various magnetic field strengths $h$. The red and blue curves represent the spectral weight evolution for the parallel (A) and antiparallel (B) sublattices, respectively. As $h$ increases, the gap in the antiparallel sublattice (B, blue) fills up faster than for the A sublattice (red), reflecting the competition induced by antiferomagnetic coupling and the Zeeman energy.
    }
    \label{fig:DOS_H}
\end{figure*}

We now turn our attention to the analysis of the system's behavior as a function of the external magnetic field strength. Before proceeding, it is important to note that the values of \( h \) considered in our simulations are relatively large when translated into physical units. For example, assuming a representative hopping amplitude of \( t = 0.2\,\mathrm{eV} \), a field value of \( h = 0.08 \) corresponds to a Zeeman energy of \(\sim0.05\,\mathrm{eV} \), which maps onto a physical magnetic field of roughly \( 500\,\mathrm{T} \). This is at least an order of magnitude higher than the typical range accessible in laboratory experiments. Nevertheless, such large field values are instructive in the theoretical context, as they enhance the visibility of the underlying mechanisms — such as the suppression of antiferromagnetic order, the emergence of sublattice asymmetries, and the resulting changes in magnetotransport properties — within the DMFT framework.

As in the temperature study, we begin by presenting the magnetization results in Figure~\ref{fig:magnetization_h}. At low magnetic fields, the system exhibits a clear AF order, characterized by nearly equal and opposite magnetizations on the two sublattices ($m_A \approx -m_B$). As the magnetic field strength increases, this symmetry breaks: the parallel sublattice ($m_A$) aligns more strongly with the field, while the antiparallel sublattice ($m_B$) weakens due to competition between AF coupling and Zeeman energy. At higher fields, beyond a critical threshold $h_c$, the AF order is completely suppressed and the system transitions into a polarized PM state. This critical value $h_c$ is temperature dependent: it increases at lower temperatures, where AF order is more robust, and decreases as thermal fluctuations assist the suppression. The field $h^*$ discussed in the temperature section corresponds to the limiting value $h_c$ as \( T \to 0 \). 

At low temperature, below\(~ T \sim 0.03\), we found a magnetic behavior reminiscent of a first-order transition. While previous studies have reported metamagnetic transitions within the paramagnetic phase, such as the correlated metal--insulator jump found by Laloux, Georges, and Krauth~\cite{Laloux1994}, our results suggest that a similar first-order mechanism can also emerge within the antiferromagnetic phase, driven by an abrupt field-induced suppression of Néel order. Those results will be reported elsewhere. 


Figure~\ref{fig:transport_H} presents the magnetic-field dependence of the transport properties. Panels (a) and (b) show the conductivity $\sigma(h)$ and resistivity $\rho(h)$, respectively. As the field increases, the system evolves from an insulating AF phase into a metallic PM phase. 
At low fields, competition between AF coupling and the Zeeman energy narrows the gap, enhancing conductivity and resulting in a "bad insulating" regime. At high fields, the increased magnetization favors Pauli exclusion, reducing the mobility of charge carriers and leading to a rise in resistivity. 

In panel (c), the scattering rates reflect the breakdown of sublattice symmetry under increasing field. Fluctuations in the antiparallel sublattice \( B \) grow significantly stronger than those in \( A \). This imbalance is reversed at high fields in the PM phase, where scattering rates decrease due to the reduced mobility. 
Above the Néel temperature ($T_N\sim 0.092$) the PM behavior dominates across all field strengths, as illustrated for $T=0.10$.

Panel (d) shows the results obtained for the MR which exhibits a clear quadratic dependence on the field. As seen in figure~\ref{fig:transport_H}, above the $T_N$, the MR is positive and increases with the applied field. In the metallic PM phase, stronger magnetic fields enhance magnetic order, which increases Pauli's exclusion effects between conducting electrons, thereby raising resistivity. Below the $T_N$, at intermediate temperatures, the MR is initially negative and decreases with small fields but abruptly changes concavity at the critical fields $h_c$, eventually becoming positive. This abrupt change in MR gradually smooths out at lower temperatures.

A remarkably similar MR profile has been reported for the topological antiferromagnetic insulator \ce{EuSn2As2} (see Fig. 3(b) in~\cite{Chen2020}). Although the mechanisms in this material are likely more complex than those in our study, notable parallels emerge. Just as in our model, MR is negative in the AF phase and turns positive upon suppression of magnetic order. Above \( T_N \), positive MR is also recovered. These features suggest that the field-driven suppression of AF order may constitute a generic route to MR sign reversal in a broad class of correlated systems.

Finally, we analyze the evolution of the spectral weight. Figure~\ref{fig:DOS_H} shows the spectral function \( A(\omega) \) for both sublattices at \( T = 0.06 \). At low fields, a clear spectral gap is present, consistent with the AF insulating state. As \( h \) increases, the Zeeman energy lifts the sublattice degeneracy, and stronger fluctuations in the antiparallel sublattice \( B \) broaden its spectrum more rapidly. Beyond the critical field \( h_c \), both sublattices recover symmetry in the PM regime, and a quasiparticle peak emerges at \( \omega = 0 \), signaling the onset of metallic behavior.


\section{Conclusions}

In this work, we have investigated the magnetotransport properties of the half-filled antiferromagnetic (AF) one-band Hubbard model in the presence of an external magnetic field, using single-site dynamical mean-field theory (DMFT). Our analysis focused on the mechanisms underlying magnetoresistivity (MR), with particular attention to its dependence on temperature and magnetic field strength.

In both our model and in giant magnetoresistance (GMR) systems, the transition from an antiparallel to a parallel magnetic configuration leads to a reduction in resistivity. However, while GMR is driven by spin-dependent scattering at interfaces, in our model the MR arises from sublattice-dependent scattering processes, where resistivity is closely linked to the local magnetization of each sublattice rather than to spin channels alone. Despite this fundamental difference, both mechanisms highlight the profound influence of magnetic order on electronic transport and illustrate how qualitatively similar MR behavior can emerge from distinct microscopic origins.


Our findings also establish a strong connection with experimental results in a variety of correlated electron systems. Similar to the Ti-doped compound \ce{Ca3Ru2O7} \cite{Zhu2016}, we observe a field-driven collapse of the Mott gap resulting from the suppression of the AF order. This suppression is reflected in the MR as a sign change, a phenomenon also observed in \ce{V2O3} samples \cite{Trastoy2020}. Beyond this, we identify other systems with MR behaviors governed by different microscopic mechanisms yet exhibiting strikingly similar MR profiles. Examples include the AF heavy-fermion compounds \ce{CePtIn4} and \ce{YbRh2Si2} \cite{Das2019, Knapp2025}, which display analogous field-dependent transitions. Particularly notable are the mixed-valent vanadate \ce{PbV6O11} \cite{Maignan2010}, which shares a remarkably similar temperature-dependent MR profile despite undergoing a ferromagnetic-to-paramagnetic transition, and the AF topological insulator \ce{EuSn2As2} \cite{Chen2020}, whose MR response to magnetic fields closely resembles that of our model.

The insights gained from this study highlight the fundamental role of magnetic-field-induced fluctuations and strong correlations in governing MR in AF systems. By analyzing the simplest model that captures AF order in a correlated electronic environment, we provide a baseline understanding of MR behavior that is directly relevant to more complex systems. This work thus serves as a foundation for future extensions—such as the inclusion of multiorbital interactions, spin-orbit coupling, and topological effects—which are expected to further deepen our understanding of magnetic transport in real materials. In this regard, our results may serve as a useful benchmark for more realistic band-structure+DMFT calculations aimed at the theoretical study of specific compounds.

Finally, we note that our analysis has considered a magnetic field aligned with the Néel axis. For small deviations from this alignment, the qualitative features of the magnetoresistive response are expected to remain robust, although quantitative aspects such as critical temperatures or field thresholds may vary. In contrast, a strictly perpendicular field would generate transverse magnetization components and spin canting~\cite{Brown2017}, requiring a more general self-consistent ansatz and potentially leading to non-diagonal hybridization functions. Such configurations pose technical challenges for single-site DMFT and remain an interesting subject for future investigation.

\section{Acknowledgments} The authors acknowledge support from UBACyT (Grant No. 20020170100284BA) and Agencia Nacional de Promoción de la Investigación, el Desarrollo Tecnológico y la Innovación (Grant No. PICT-2018-04536).

\appendix

\section{Derivation of the expression for the conductivity} \label{conduc}

The expression for the conductivity in the context of DMFT, for a bipartite lattice with antiferromagnetic symmetry, has been derived in Ref.~\cite{Pruschke2003}. Following that approach, we present here the general derivation for such a lattice structure.

As discussed in Sec.~\ref{model}, it is convenient to introduce the creation operators $a^\dagger$ and $b^\dagger$, which act on sublattices $A$ and $B$, respectively. The Hamiltonian can then be written as,
\begin{multline}
H = -t\sum_{\langle ij \rangle, \sigma}\left( a^\dagger_{i \sigma} b_{j \sigma} + b^\dagger_{i \sigma} a_{j \sigma} \right)+
 \\+ U\sum_i n_{i\uparrow}n_{i\downarrow} -\sum_{i, \sigma} \mu n_{i \sigma}.
\end{multline}
where $t$ is the hopping amplitude, $U$ is the on-site Coulomb repulsion, and $\mu$ is the chemical potential.

To introduce the coupling to an external electromagnetic field, we apply the Peierls' substitution. The current operator $j_\alpha$ for $\alpha = x, y, z$ is then given by,
\begin{multline} 
j_\alpha=j_\alpha^p+j_\alpha^d=
\sum_{\mathbf{k},\sigma} \Psi^\dagger_{\mathbf{k},\sigma}
\left(\begin{matrix}
        0 & \frac{\partial \ve_\mathbf{k}}{\partial \mathbf{k}_\alpha} \\
        \frac{\partial \ve_\mathbf{k}}{\partial \mathbf{k}_\alpha} & 0 
       \end{matrix} \right) \Psi_{\mathbf{k},\sigma} - \\
-A_\alpha \sum_{\mathbf{k},\sigma} \Psi^\dagger_{\mathbf{k},\sigma}
\left(\begin{matrix}
        0 & \frac{\partial^2 \ve_\mathbf{k}}{\partial \mathbf{k}_\alpha^2} \\ 
        \frac{\partial^2 \ve_\mathbf{k}}{\partial \mathbf{k}_\alpha^2} & 0 
       \end{matrix} \right) \Psi_{\mathbf{k},\sigma},
\end{multline}
where $A_\alpha$ denotes the Cartesian components of the vector potential, and $\Psi^\dagger_{\mathbf{k},\sigma} = \left(a^\dagger_{i \sigma}, b^\dagger_{j \sigma}\right)$ is the spinor representation. Here, $j_\alpha^p$ refers to the “paramagnetic” term, and $j_\alpha^d$ to the “diamagnetic” term.

Following the standard Kubo formulation, the optical conductivity reads,
\be
\sigma^\mathrm{op}_{\alpha\beta}(i\Omega_n)=\frac{1}{\Omega_n}\left[\Pi_{\alpha\beta}(i\Omega_n)+
\langle n_\alpha \rangle \delta_{\alpha\beta} \right],
\label{op_conduc}
\ee
where $\Pi_{\alpha\beta}(i\Omega_n) = \langle j^p_\alpha(i\Omega_n) j^p_\beta(0)\rangle$ is the current-current correlation function in bosonic Matsubara frequencies $i\Omega_n$, and $\langle n_\alpha \rangle = -\langle j_\alpha^d \rangle/A_\alpha$. The corresponding real-frequency optical conductivity is obtained by analytically continuing $i\Omega_n \to \nu + i0^+$.

The DC conductivity $\sigma$ is the real part of the optical conductivity evaluated at zero frequency, and since the last term in (\ref{op_conduc}) contribute only to the imaginary part, we will drop it hereafter.

Assuming an isotropic lattice where all directions are equivalent, the optical conductivity tensor becomes diagonal, and the elements reduce to $\sigma^\mathrm{op}_{\alpha\beta}(i\Omega_n) = \sigma^\mathrm{op}(i\Omega_n)$. Following Ref.~\cite{Pruschke2003}, we evaluate the current-current correlation function using Green’s functions, yielding,
\begin{multline}
\sigma^\mathrm{op}(i\Omega_n)=\frac{c}{\Omega_n \beta}\sum_{i\omega_n}\sum_\sigma \int\mathrm{d}\ve D(\ve)\times \\
\left[ G_\sigma^{AA}(i \omega_n+i\Omega_n,\ve)G_\sigma^{BB}(i \omega_n,\ve) + \right.\\
\left. \;\; + G_\sigma^{BB}(i \omega_n+i\Omega_n,\ve)G_\sigma^{AA}(i \omega_n,\ve) + \right. \\
\left. + 2 G_\sigma^{AB}(i \omega_n+i\Omega_n,\ve)G_\sigma^{AB}(i \omega_n,\ve) \right],
\label{op_co_G}
\end{multline}
where $c$ is a dimensional constant, $D(\ve)$ is the density of states of the hypercubic lattice, and the Green's functions [see Eq.~(\ref{inv_fullG})] are given by:
\begin{eqnarray}
G_\sigma^{\alpha \alpha}(i \omega_n,\ve) &=& \frac{\zeta_{\bar{\alpha}\sigma}}{\zeta_{A\sigma}\zeta_{B\sigma}-\ve^2} \\
G_\sigma^{\alpha \beta}(i \omega_n,\ve)=G_\sigma^{\beta \alpha}(i \omega_n,\ve) &=& \frac{\ve}{\zeta_{A\sigma}\zeta_{B\sigma}-\ve^2}.
\end{eqnarray}

When an external uniform magnetic field $h$ is applied, $\zeta_{\alpha \sigma} = i \omega_n + \mu - \sigma h-\Sigma_{\alpha \sigma}(i \omega_n)$. The field breaks the antiferromagnetic symmetry, requiring all terms in Eq.(\ref{op_co_G}) to be included. After evaluating the Matsubara summations and introducing the spectral representation,
\be
G_\sigma^{\alpha \beta}(i \omega_n,\ve)=\int \mathrm{d}\omega\frac{A^{\alpha \beta}(\omega,\ve)}{i\omega_n-\omega}
\ee
the optical conductivity becomes, 
\begin{multline}
\sigma^\mathrm{op}(i\Omega_n)=\frac{c}{\Omega_n}\sum_\sigma \int\mathrm{d}\ve \int\mathrm{d}\omega \int\mathrm{d}\omega^\prime D(\ve)\times \\
\frac{f(\omega)-f(\omega^\prime)}{\omega-\omega^\prime + i\Omega_n} 
\left[ A_\sigma^{AA}(\omega^\prime,\ve)A_\sigma^{BB}(\omega,\ve) + \right.\\
\left. \;\; + A_\sigma^{BB}(\omega^\prime,\ve)A_\sigma^{AA}(\omega,\ve) + \right. \\
\left. + 2 A_\sigma^{AB}(\omega^\prime,\ve)A_\sigma^{AB}(\omega,\ve) \right],
\label{op_co_A}
\end{multline}
where $f(\omega)$ is the Fermi function.

Finally, taking the real part of the analytically continued $\sigma^\mathrm{op}$ yields the conductivity:
\begin{multline}
\sigma(\omega)=\Re\left[\sigma^\mathrm{op}(\nu+i0^+)\right]=\pi c \sum_\sigma \int\mathrm{d}\ve \int\mathrm{d}\omega D(\ve)\times \\
\frac{f(\omega)-f(\omega+\nu)}{\nu} 
\left[ A_\sigma^{AA}(\omega +\nu,\ve)A_\sigma^{BB}(\omega,\ve) + \right.\\
\left. \;\; + A_\sigma^{BB}(\omega+\nu,\ve)A_\sigma^{AA}(\omega,\ve) + \right. \\
\left. + 2 A_\sigma^{AB}(\omega+\nu,\ve)A_\sigma^{AB}(\omega,\ve) \right].
\label{Reop_co}
\end{multline}
In the absence of a magnetic field ($h = 0$), i.e., when antiferromagnetic symmetry is restored,
$A_\sigma^{AA}(\omega,\ve) = A_{\bar{\sigma}}^{BB}(\omega,\ve)$ and
$A_\sigma^{AB}(\omega,\ve) = A_{\bar{\sigma}}^{AB}(\omega,\ve)$, and the expression of Ref. \cite{Pruschke2003} is recovered.

The DC conductivity is defined as the zero-frequency limit:
\begin{multline}
\sigma=\sigma(\nu \to 0)=2 \pi c \sum_\sigma \int\mathrm{d}\ve \int\mathrm{d}\omega D(\ve) \left(-\frac{\partial f}{\partial \omega}\right) \times \\
\left[ A_\sigma^{AA}(\omega,\ve)A_\sigma^{BB}(\omega,\ve) + \left(A_\sigma^{AB}(\omega,\ve)\right)^2 \right].
\label{conduct}
\end{multline}

\bibliography{MR_bib}

\end{document}